\documentclass[authoryear,preprint,review,12pt]{elsarticle}

\usepackage{amssymb}

\journal{Physics Letters B}

\begin{document}

\begin{frontmatter}



\title{On the dynamics of exotic matter:\\ towards creation of Perpetuum Mobile
of third kind }


\author{Pavel Ivanov}

\address{Astro Space Centre of PN Lebedev Physical Institute,
84/32 Profsoyuznaya Street, Moscow, 117810, Russia\\
Department of Applied Mathematics and Theoretical Physics, CMS,
University of Cambridge, Wilberforce Road, Cambridge, CB3 0WA, UK}

\begin{abstract}
The one-dimensional dynamics of a classical ideal 'exotic' fluid
with equation of state $p=p(\epsilon) < 0$ violating the weak
energy condition is discussed. Under certain assumptions it is
shown that the well-known Hwa-Bjorken exact solution of
one-dimensional relativistic hydrodynamics is confined within the
future/past light cone. It is also demonstrated that the total
energy of such a solution is equal to zero and that there are
regions within the light cone with negative $(-)$ and positive
$(+)$ total energies. For certain equations of state there is a
continuous energy transfer from the $(-)$-regions to the
$(+)$-regions resulting in indefinite growth of energy in the
$(+)$ regions with time, which may be interpreted as action of a
specific 'Perpetuum Mobile' (Perpetuum Motion). It is speculated
that if it is possible to construct a three-dimensional
non-stationary flow of an exotic fluid having a finite negative
value of energy such a situation would also occur.  Such a flow
may continuously transfer positive energy to gravitational waves,
resulting in a runaway. It is conjectured that theories plagued by
such solutions should be discarded as inherently unstable.
\end{abstract}


\end{frontmatter}


\section{Introduction}

If physical laws do not prohibit the presence of exotic matter
violating the weak energy condition$^{1}$\footnotetext[1]{Let us
remind that the weak energy condition is said to be violated for a
matter having the stress-energy tensor $T_{\mu\nu}$ if there is
such a time-like future directed vector field $t^{\mu}$ that the
inequality $T_{\mu\nu}t^{\mu}t^{\nu} < 0$ holds in some region of
space-time.} and having some other certain properties many
exciting possibilities arise. For example, solutions of the
Einstein equations coupled to an exotic matter include wormholes,
time machines (e.g. \cite{1}) and cosmological models with energy
density of the Universe growing with time (see, e.g. \cite{2} for
a review and discussion of cosmological consequences) leading to the
so-called cosmological Doomsday, see e.g. \cite{2b}. Since
theories incorporating an exotic matter may lead to
counterintuitive and, possibly, physically inconsistent effects it
appears to be important to invoke different thought experiments,
which could clarify self-consistency of such theories. Here we
discuss such an experiment and explicitly show that in a class of
models containing an exotic matter of a certain kind there could
be ever expanding with time separated regions of space having
positive and negative total energies and that absolute values of
the energies in these regions could grow indefinitely with time
while the energy of the whole physical system is conserved. This
is based on the property of the exotic matter to have negative
energy density measured by observers being at rest with respect to
some Lorentz frame and, accordingly, given by the $(tt)-$component
of the stress energy tensor, $T^{tt}$, provided that there are
sufficiently large fluid velocities with respect to this frame. We
also speculate that in a more advanced variant of our model there
could be an isolated region of space filled by an exotic matter
with its total energy indefinitely decreasing with time due to
processes of interaction with some other conventional physical
fields carrying positive energy. One of such processes could be
emission of gravitational waves. If conditions for emission of
gravitational waves are always fulfilled in the course of
evolution positive energy is continually carried away from the
region, which results in a runaway. In this respect it is
appropriate to mention that the known results on positiveness of
mass in General Relativity are not valid for the matter violating
the weak energy condition, see e.g. \cite{3} and the energy of an
isolated region could evolve from positive to negative values.
Although such a situation resembles the action of a Perpetuum
Mobile of second kind, where heat transfer from a colder part of an
isolated system to a hotter part occurs, the notion of
'temperature' looks ambiguous in our case and, therefore, because
of the lack of notation, we refer to this hypothetical effect as a
'Perpetuum Mobile of third kind'. In many applications a
phenomenological description of the exotic matter assumes that the
matter dynamics can be described by a hydrodynamical model with
the stress-energy tensor of an ideal Pascal fluid, which can be
fully specified by its equation of state, $p=p(\epsilon )$, where
$p$ is the pressure and $\epsilon$ is the comoving energy density,
which is assumed to be positive below.
In this case, violation of the weak energy condition can be
reformulated as a requirement that the pressure is negative with
its modulus exceeding the value of the comoving energy density. We
deal hereafter only with simplest models of hydrodynamical type
and neglect the effects of General Relativity and interactions
with other physical fields. Therefore, in the model explicitly
considered in the text the transfer of energy from one region of
space to another, in particular, between the regions having total
energies of different signs, is provided by hydrodynamical
effects. However, as we have mentioned above, it seems reasonable
to suppose that an analogous runaway effect may happen in a more
realistic situation, where e.g. energy is continually carried away
from a spacial region having negative total energy by
gravitational waves$^2$\footnotetext[2]{ Note that one should
distinguish between the standard hydrodynamical instabilities e.g.
the ones operating in a fluid having a negative square of speed of
sound and the instability related to violation of the weak energy
condition. While the former could lead to a highly irregular
non-linear dynamics of the system they cannot themselves result in
formation of spacial regions having a negative total energy, and,
a runaway of the kind discussed in this Paper, see also the
footnote in the next Section.

Let us also stress that the runaway effect is
different from the well-known instabilities of linear
modes of stationary hydrodynamical flows having a
negative energy difference with respect to a stationary configuration
like the Chandrasekhar-Friedman-Schutz
instability \cite{3a}, \cite{3b} or the instability operating in
shear flows, see e.g. \cite{3c} and references therein.
Although the linear instabilities can lead to a descrease of
energy of the main flow they cannot bring a system non-violating the weak
energy condition to a state with a negative total energy. Therefore,
in the latter case the runaway effect of the type we consider in this Paper
is not possible.}. Taking into account that gravitational
interaction is universal, the proposed 'Perpetuum Mobile of third
kind' may represent a difficulty in theories, where it emerges. We
believe that such theories are inherently unstable and must,
therefore, be discarded.

It is important to note, however, that in a general scenario the region emitting
gravitational waves could have a non-linear three-dimensional dynamics.
Explicit solutions of this kind may be quite difficult to obtain due to
severe technical problems.

\section{ Basic definitions and equations}

Let us discuss a one-dimensional planar relativistic flow of an
ideal fluid with a baratropic equation of state
\begin{equation}
p=p(\epsilon), \label{1}
\end{equation} where $p$ is the pressure
and $\epsilon $ is the comoving energy density. As has been
mentioned in Introduction, we are going to consider later in the
text the case of an exotic fluid, where the pressure is negative
and the weak energy condition is violated. For a baratropic fluid
this leads to:
\begin{equation}
\sigma \equiv -p > \epsilon, \label{2}
\end{equation}
 where we introduce the
negative of pressure $p$, $\sigma = -p$. Since only
one-dimensional flows will be considered, we can also apply our
analysis to a situation, where a fluid has an anisotropic stress
tensor. Say, we can assume only one of its components to be
non-negative and proportional to $\delta (y,z) $, where $y$ and
$z$ are the Minkowski spacial coordinates corresponding to
directions perpendicular to the direction of motion. In this case,
equations of motion will describe dynamics of a straight string
consisting of exotic matter.

Equations of motion may be written in a divergent form reflecting
the laws of conservation of energy and momentum
\begin{equation}
T^{tt}_{,t}+T^{tx}_{,x}=0, \quad T^{tx}_{,t}+T^{xx}_{,x}=0,
\label{4}
\end{equation} where $(t,x)$
are the standard Minkowski coordinates, comma stands for
differentiation, and $T^{ij}$ are the corresponding components of
the stress-energy tensor:
\begin{equation}
T^{tt}=\gamma^{2}(\epsilon + v^{2}p), \quad
T^{tx}=\gamma^{2}v(\epsilon + p), \quad T^{xx}=\gamma^{2}(p +
v^{2}\epsilon), \label{5}
\end{equation} where $v$ is the three-velocity and
$\gamma ={1\over \sqrt {1-v^{2}}}$.

\section{The Hwa-Bjorken solution and the Milne coordinates}

As has been first shown by Hwa \cite{4} and later by Bjorken
\cite{5}, the set of equations (\ref{4}) has an especially simple
'acceleration-free' solution valid for a fluid having an arbitrary
equation of state. In this solution velocity of any fluid element
conserves along the path of the fluid element and the velocity
field has a very simple form
\begin{equation}
v=x/t\equiv \xi. \label{7}
\end{equation}
For a baratropic fluid
the distribution of energy is given by another simple implicit
relation
\begin{equation} \tau = \exp
\lbrace{-\int^{\epsilon}_{\epsilon_{*}}{d\epsilon^{'} \over
\epsilon^{'} + p(\epsilon^{'})}}\rbrace =\exp \lbrace
-{\int^{\epsilon_{*}}_{\epsilon } {d\epsilon^{'} \over \sigma
(\epsilon^{'}) -\epsilon^{'}}}\rbrace, \label{8}
\end{equation} where
\begin{equation} \tau =\sqrt {t^{2}-x^{2}}, \label{9}
\end{equation}
and $\epsilon_{*}$ is a constant of integration.

Obviously, equations (\ref{8}-\ref{9}) are defined only inside the
future/past light cone,  $|\tau| > |x|$, in an effective
two-dimensional Minkowski space described by the metric
\begin{equation} ds^{2}=dt^{2}-dx^{2}.
\label{10}
\end{equation}
The analytic continuation of the solution on the right/left
Rindler wedge $|\tau| < |x|$ is straightforward.

Although the two-dimensional Minkowski space appears naturally due
to one-dimensional character of the problem let us remind that the
problem  is defined in four-dimensional Minkowski space.
Therefore, for the problem with Pascal pressure, where it is
assumed that all variables do not depend on the coordinates
$(y,z)$ perpendicular to $x$, it is better to say that from the
four-dimensional point of view the condition $|\tau|=|x|$
determines four-dimensional ``light wedges'' since it does not
depend on directions perpendicular to the direction of motion.

The energy density $\epsilon $ is equal to zero on the light cone
$|\tau|= |x|$ if and only if the integrals in the exponents in
equation (\ref{8}) are positive and diverge when $\epsilon
\rightarrow 0$. Accordingly, in this case, the condition (\ref{2})
must be satisfied. Additionally, in order to make the integrals
divergent we must have
\begin{equation} \sigma - \epsilon \le
|O(\epsilon)| \label{11}
\end{equation}
when $\epsilon \rightarrow 0$. Provided that condition (\ref{11})
is valid the solution may be considered as confined within the
future/past light cone. For simplicity we are going to consider
only this case in our analytical calculations later on$^3$
\footnotetext[3]{In this case, the corresponding hydrodynamical
models are unstable with respect to growth of small perturbations.
One may proceed, however, either assuming that these models are
valid only for the considered types of hydrodynamical flows or
considering them as effective models invalid for sufficiently
large perturbation wavenumbers. In any case, we expect that our
main conclusions do not depend on whether the considered models
are hydrodynamically unstable or not.}.

The solution (\ref{7}-\ref{8}) has a self-evident form in the
Milne coordinates $(\tau, y)$, where the time $\tau $ is defined by
equation (\ref{9}) and we introduce the rapidity
\begin{equation}
y=\ln \sqrt {{1+\xi \over 1- \xi}}  \label{12}
\end{equation} as a new spacial coordinate. In these
coordinates the metric (\ref{10}) has the form
\begin{equation}
ds^{2}=d\tau^{2}-\tau^{2}dy^{2}. \label{13}
\end{equation}
Although the metric (\ref{13}) may be understood as describing an
expanding one-dimensional spacially-uniform universe with the
scale factor $a(\tau)=\tau$, obviously it corresponds to the same
flat Minkowski space since it is obtained from the metric
(\ref{10}) by the coordinate transformation (\ref{9}), (\ref{12}),
see e.g. \cite{9a} for an additional discussion of the Milne
coordinates. Taking into account that the transformation law
between the velocity $v$ defined with respect to Minkowski
coordinates $(t,x)$ and the peculiar velocity $\bar v$ defined
with respect to the orthonormal frame associated with the Milne
coordinates is determined by the relation
\begin{equation} v= {\xi + \bar v \over 1 + \xi \bar v},
\label{14}
\end{equation} it is clear that the solution (\ref{7}-\ref{8}) is
simply the spacially-uniform solution in the Milne coordinates
with the peculiar velocity $\bar v=0$ . In particular, equation
(\ref{8}) immediately follows from the first law of thermodynamics
written for an adiabatic expansion of a fluid having distribution
of its thermodynamical variables uniform with respect to the
coordinate $y$:
\begin{equation} {dV \over V} = -{d\epsilon \over \epsilon + p},
\label{15}
\end{equation}
where $V \propto \tau$ is a comoving volume.

\section{Properties of the solution}

\subsection{The energy integral}

Provided that the condition (\ref{11}) is valid we can assume that
the energy density and, accordingly, the components of the
stress-energy tensor are different from zero only within the
future/past light cone. For simplicity, let us consider only the
future light cone implying that $t > 0$ from now on. In this case,
equations (\ref{4}) yield that the total energy of the flow
\begin{equation}
E=\int^{t}_{-t}dx T^{tt} = t\int^{1}_{-1}d\xi T^{tt} \label{16}
\end{equation} does not depend on time $t$.

Let us show that this integral is precisely equal to zero for an
exotic baratropic fluid satisfying the condition (\ref{11}).
Taking into account that the distribution of energy density and
pressure are even functions and the velocity distribution is an
odd function of the spacial coordinate $x$, respectively, it
suffices to prove that the quantity \begin{equation} {\cal E} =
\int^{1}_{0}d\xi T^{tt}= \int^{1}_{0}{d\xi \over
(1-\xi^{2})}(\epsilon - \xi^{2}\sigma ) \label{17}
\end{equation} is equal to zero. From
equation (\ref{8}) it follows that
\begin{equation}
\sigma =\tau {d\epsilon \over d\tau } + \epsilon \label{18}
\end{equation} and,
therefore, we have \begin{equation} {\cal E} = \int^{1}_{0}d\xi
(\epsilon - {\xi^{2}\tau \over (1-\xi^{2})} {d\epsilon \over d\tau
}). \label{19}
\end{equation}

Now we change the integration variable from $\xi $ to $\tau $
keeping the value of $t$ fixed. Taking into account that
\begin{equation} d\xi =-{\tau \over t \sqrt {t^{2} -
\tau^{2}}}d\tau, \label{20}
\end{equation}  \begin{equation} \xi = {1\over
t}\sqrt {t^{2} -\tau^{2}}, \quad \sqrt {1-\xi^{2}}={\tau \over t},
\label{21}
\end{equation} we obtain \begin{equation} {\cal E} = {1\over t}\int
\sqrt {t^{2} -\tau^{2}}d\epsilon  - {1\over t}\int {\epsilon \tau
d\tau \over \sqrt {t^{2} -\tau^{2}}}, \label{22}
\end{equation} where the values of $\tau $ and $\epsilon $
corresponding to $\xi=0, 1$ are omitted. Integrating by parts the
first integral in (\ref{22}) and taking into account that the
boundary terms are equal to zero for the solution
(\ref{7}-\ref{8}) satisfying the condition (\ref{11}), we obtain
\begin{equation}
{\cal E} = -{1\over t}\int  ({\tau \over \sqrt {t^{2} -\tau^{2}}}
+{d\over d\tau }\sqrt {t^{2} -\tau^{2}})\epsilon d\tau =0.
\label{23}
\end{equation}

\subsection{Lorentz invariance and vacuum-like nature}

It is easy to see that solution (\ref{7}-\ref{8}) has the same
form in all coordinate systems connected by the Lorentz
transformations: $(t, x) \rightarrow (t',x')$. Indeed, as follows
from equation (\ref{9}) the time $\tau $ is invariant under the
Lorentz transformations. Therefore, equation (\ref{8}) contains
only invariant quantities and is the same in all Lorentz frames.
It is also evident that when the Lorentz transformations are
considered the quantity $\xi = x/t $ and the three-velocity $v$
are transformed in the same way. Therefore, from equation
(\ref{7}) it follows that the same equation is valid for the
transformed quantities.

It is clear that the total energy and momentum of the flow are
equal to zero in all Lorentz frames. This is frequently considered
as being the definition of vacuum solutions in different
theoretical schemes. Thus, one may state that solution
(\ref{7}-\ref{8}) plays a role of a non-trivial vacuum solution
for the exotic fluids satisfying (\ref{11}).

\subsection{A hypothetical model of Perpetuum Mobile}

As has been mentioned in Introduction, the very possibility of
existence of solutions having negative/zero total energy is
determined by the fact that for equations of state violating the
weak energy condition the energy density determined with respect
to a fixed Lorentz frame
\begin{equation}T^{tt} \propto \epsilon -v^{2}\sigma \label{24}\end{equation}
can be negative provided that the fluid velocity is sufficiently
large,
\begin{equation}v > v_{crit}=\sqrt{{\epsilon \over \sigma}}. \label{25}\end{equation}
In the case of our solution the space bounded by the light cone
condition $|x| < t$ is divided into a set of regions having
opposite signs of  $T^{tt}$ and, accordingly, different signs of
the total energy. In what follows let us refer to the regions with
$T^{tt} > 0$ ($T^{tt}
< 0$) as $(+)$-regions ($(-)$-regions). The coordinates of
boundaries between the $(+)$ and $(-)$-regions, $x_{crit}$, can be
found from the condition $v=v_{crit}$ and, respectively, from the
implicit equation
\begin{equation}x_{crit}=
t\sqrt{{\epsilon(\tau_{crit})\over
\sigma(\tau_{crit})}},\label{26}\end{equation} where
$\tau_{crit}=\sqrt{t^{2}-x_{crit}^{2}}$. In general, equation
(\ref{26}) could have several roots on the interval $0< x <t$ with
values depending on equation of state. However, for a reasonable
equation of state with sufficiently smooth dependence of $\sigma $
on $\epsilon $ there must be a $(-)$-region adjacent to the light
cone boundary $x=t$ and a $(+)$-region close to the point $x=0$.
The total energy of the region adjacent to the light cone,
$E_{-}$, can be easily calculated from equations (\ref{16}),
(\ref{17}), and (\ref{22}), where we take into account that after
integration by parts of equation (\ref{22}) only the boundary term
at $x=x_{crit}$ contributes to the integral:
\begin{equation} E_{-}=\int^{t}_{x_{crit}}T^{tt}dx=-x_{crit}\epsilon, \label{27}\end{equation}
where $x_{crit}$ denotes the largest root of equation (\ref{26})
in the interval $0< x <t$ from now on. Taking into account the
fact that the total energy is zero and  using the symmetry of the
problem, we see that the energy in the interval $0<x<x_{crit}$,
$E_{+}=-E_{-}$, and therefore
\begin{equation} E_{+}=\int_{0}^{x_{crit}}T^{tt}dx={t\epsilon^{3/2}\over \sigma^{1/2}}
={\epsilon^{3/2}\over \sqrt {\sigma - \epsilon}}
\exp{(\int^{\epsilon} {d\epsilon^{'}\over \sigma^{'}
-\epsilon^{'}})}, \label{28}\end{equation} where we use equations
(\ref{8}), (\ref{26}), (\ref{27}) and all quantities are assumed
to be evaluated along the world line determined by equation
(\ref{26}). It is instructive to calculate the time derivative of
$E_{+}$ differentiating the integral in (\ref{28}) on time and
using equations (\ref{4}) and (\ref{25}) to obtain:
\begin{equation}
\dot E_{+}=\sigma v=\sqrt{\sigma \epsilon}. \label{29}
\end{equation}
From equation (\ref{29}) it follows that the energy of the region
$0<x<x_{crit}$ constantly grows with time due to energy flow from
the (-)-region $x_{crit} < x < t$. In principal, there could be
two possibilities for the asymptotic behaviour of $E_{+}$ in the
limit $t\rightarrow \infty $ depending on form of the equation of
state: 1) there could be a finite asymptotic value of $E_{+}$, and
2) a finite asymptotic value could be absent and the energy
$E_{+}$ could infinitely grow with time. The latter case
represents a specific instability, where there is infinite growth
of energy in one region of space and infinite decrease of energy
in the other region. Let us suppose that physical laws do not
prohibit existence of the hydrodynamical systems violating the
weak energy condition and having solutions of this type. Assuming
that such a system could be made by an advanced civilization, which
could also ensure that utilization of energy released in the
region $0<x<x_{crit}$ does not significantly perturb the solution
this could provide an infinite source of energy. Therefore, this
type of solution may be classified as a hypothetical 'Perpetuum
Mobile'. Such a solution is discussed below.

\subsection{An explicit example}

Let us specify the equation of state and consider the simplest
possible case of linear dependence of $\sigma $ on $\epsilon $ :
\begin{equation} \sigma =(1+\alpha) \epsilon,
\label{30}
\end{equation}
where the parameter $\alpha > 0$.$^4$ \footnotetext[4]{Let us
stress that there is no well defined limit to the case of the
'standard' vacuum equation of state $\alpha =0$ (i.e. $p=-\epsilon
$) and that all our conclusions are not applicable to this case.}
In this case integration of equation (\ref{8}) gives
\begin{equation}
\epsilon=C\tau^{\alpha}, \label{31}
\end{equation}
where the parameter $C > 0$. The critical velocity
$v_{crit}={1\over \sqrt {1+\alpha}}$ and, accordingly
\footnotetext[5]{Note that in this case it is obvious that the
hypersurfaces separating the $(+)$ and $(-)$ regions are
time-like.},
\begin{equation}
x_{crit}= {t\over \sqrt {1+\alpha}}. \label{32}
\end{equation}
From equations (\ref{9}), (\ref{26}), (\ref{29}), (\ref{30}) and
(\ref{31}) we obtain
\begin{equation}
E_{+}={C \alpha^{\alpha/2}\over
({1+\alpha})^{(1+\alpha)/2}}t^{1+\alpha} \label{33}
\end{equation}
From equation (\ref{33}) it follows that in the case of the linear
equation of state there is one $(+)$-region and one $(-)$-region
in the range $0 < x < t$. The energy of the $(+)$-region
($(-)$-region) grows (decreases) indefinitely. Therefore, the
simplest linear equation of state determines solution, which may be
classified as the 'Perpetuum Mobile'.

\section{Discussion}

When the weak energy condition is violated the Hwa-Bjorken
solution is likely to be not the unique one having the total
energy $E$ equal to zero. Say, in the model with the linear
equation of state (\ref{30}) it is easy to find a family of
self-similar solutions, where the velocity $v$ is a function of the
self-similar variable $\xi$, $v=f_{1}(\xi)$, and the energy
density $\epsilon $ has the form:
\begin{equation}\epsilon=t^{\beta}f_{2}(\xi).\label{34}
\end{equation}
Substituting these expressions in equations (\ref{4}) and assuming
that the equation of state is given by (\ref{30}) one can easily
get two ordinary differential equations for the functions $f_{1}$
and $f_{2}$. In this case we have
\begin{equation} E=\int^{+\infty}_{-\infty} dxT^{tt}=t^{\beta +1}
\int^{+\infty}_{-\infty}d\xi f_{3}(\xi),    \label{35}
\end{equation}
where $f_{3}$ is expressed through $f_{1}$ and $f_{2}$. It is
assumed that the energy density is either equal to zero when $\xi
$ is sufficiently larger or tends to zero with increase of $\xi $
fast enough to make the integral convergent. When this condition
is fulfilled and $\beta \ne -1$ the energy $E$ is equal to zero.
Indeed, the energy must not depend on time. On the other hand it
is seen from (\ref{35}) that the energy is proportional to
$t^{\beta+1}$. This means that the integral
$\int^{+\infty}_{-\infty}d\xi f_{3}(\xi)$ must be equal to zero
provided that $\beta \ne -1$.

A more difficult and interesting problem would be to construct an
explicit solution having a negative value of the total energy. In
this case one should invoke more sophisticated methods of
one-dimensional relativistic hydrodynamics such as e.g. the
hodograph method introduced by Khalatnikov \cite{6} (see also
\cite{7}, \cite{8}). In the case of the linear equation of state
(\ref{30}) it would also be interesting to exploit the formalism
developed in Ref.  \cite{9}, where the set of equations (\ref{4})
is reduced to a single equation, which can be analysed for new
solutions.

Another approach to the problem consists in use of numerical
methods. It seems that numerical methods are more suitable for
hydrodynamically stable models e.g. based on the 'Chaplygin type'
equation of state $p=-\epsilon_{*}^{2}/\epsilon$, where
$\epsilon_{*}$ is a constant. In framework of numerical methods
one can consider a hydrodynamical motion with fixed boundary
conditions on a fixed spacial interval, e.g. $v=0$ when $x=0$ and
when $x=x_{1}$, where $x_{1}$ is fixed. The total energy of the
motion determined by initial conditions can, in principal, be
either negative or zero for fluids violating the weak energy
condition.

An interesting development of these studies would be to consider a
model, where the exotic matter having anisotropic tension is
concentrated on a straight one-dimensional line in three
dimensional space. One-dimensional motion excited on the line
could produce gravitational waves carrying away positive energy
from the system. Accordingly, the energy of the motion could
decrease indefinitely provided that generation of gravitational
waves persists in the course of evolution of the system. Being
extrapolated on realistic three-dimensional motions this effect
could be dangerous for models of exotic matter, which can be
effectively described in the hydrodynamical approximation. The
energy of such motions could decrease indefinitely, resulting in a
runaway. In this case, the corresponding models should be
discarded.

It would be also interesting to look for a quantum model having
the properties disscussed in the Paper. Note that in quantum case
some exotic properties of behaviour of field systems  may be
expected even when they have a "normal" classical limit. Say, as
was discussed e.g. in Ref \cite{15a}, in the Milne universe vacuum
expectation value of the comoving energy density can be negative
even for the simplest model of a non-interacting scalar field with
sufficiently small mass, in a certain vacuum state.

At the end we would like to point out that although the runaway
process related to emission of gravitational waves may be
technically difficult to construct for the exotic matter with
positive comoving energy density, it can be easily constructed for
even more exotic 'ghost' matter having a negative value of the
energy density in all frames implemented by physical bodies and
clocks. For example, we can make use of the model of rotating
relativistic string with two monopoles at its ends emitting weak
gravitational waves, see Ref. \cite{10}. It is sufficient to
change the sign of the Lagrangians describing the string and the
monopoles while keeping the sign of the gravitation part of the
action fixed to convert this model to a model of a 'ghost' matter
interacting with gravity. It is clear that neither the string
dynamics nor conditions of emission of gravitational waves are
significantly affected by this procedure. Such a  model describes
a finite length string with its length ever increasing with time
thus making the total energy of the string-monopoles system ever
decreasing. The positive energy carried away by gravitational
waves may be exploited by an advanced civilization able to
constract such a device.

I am grateful to V. I. Dokuchaev, I. D. Novikov, V. N. Lukash, G.
W. Gibbons, J. C. B. Papaloizou and Vovik Strokov for useful
remarks. It is my sincere pleasure to thank D. A. Kompaneets for
many important discussions and comments.  This work has been
supported in part by RFBR grants $07-02-00886-$a and
$08-02-00159-$a and by the governmental grant NSh-2469.2008.2.




\begin{thebibliography}{00}


\bibitem{1}
    M. S. Morris, K. S. Thorne, U. Yurtsever,
    Phys. Rev. Lett.,  61, 1446 (1988)

\bibitem{2}
    E. J. Copeland, M. Sami, S. Tsujikawa,
    Int. J. of Mod. Phys. D, 15, 1753 (2006)

\bibitem{2b} R. R. Caldwell, M. Kamionkowski, N. N. Weinberg,
Phys. Rev. Lett, 91, 071301 (2003)

\bibitem{3} R. Schoen, S. T. Yau, Phys. Rev. Lett. 43, 1457
(1979); R. Schoen, S. T. Yau, {\it ibid} 48, 369 (1982); G. T.
Horowitz, M. J. Perry, {\it ibid} 48, 371 (1982)

\bibitem{3a} S. Chandrasekhar, Phys. Rev. Lett., 24, 611 (1970)

\bibitem{3b} J. L. Friedman, B. F. Schutz, Ap. J., 222, 281 (1978)

\bibitem{3c} W. Glatzel, Reviews in Modern Astronomy, 4, 104 (1991)

\bibitem{4} R. C. Hwa, Phys. Rev. D 10, 2260 (1974)

\bibitem{5} J. D. Bjorken, Phys. Rev. D 27, 140 (1983)

\bibitem{9a} N. D. Birrell, P. C. W. Davies, ``Quantum Field in Curved
Space'', Cambridge University Press, Cambridge, UK (1982)

\bibitem{6} I. M. Khalatnikov, Zh. Eksp. Teor. Fiz. 27, 529 (1954)

\bibitem{7} E. P. T. Liang, Astrophys. J. 211, 361 (1977)

\bibitem{8} P. Carbonaro, Phys. Rev. E 56, 2896 (1997)

\bibitem{9} A. Bialas, R. A. Janik, R. Peschanski, Phys. Rev. C 76, 054901
(2007)

\bibitem{15a} T. S. Bunch, S. M. Christensen, S. A. Fulling, Phys. Rev. D 18, 4435
(1978)

\bibitem{10} X. Martin, A. Vilenkin, Phys. Rev. D 55, 6054 (1997)




\end{thebibliography}
\end{document}